\documentclass[a4paper,11pt]{article}

\usepackage{amsfonts}
\usepackage[dvips]{color,graphicx}
\usepackage{graphicx,amssymb,amsmath,bm,latexsym}
\usepackage{stmaryrd}
\usepackage{bm}

\definecolor{red  }{rgb}{1,0,0}
\definecolor{blue }{rgb}{0,0,1}
\definecolor{green}{rgb}{0,1,0}

\newcommand{\vs}[1]{\vspace{#1 mm}}

\usepackage{graphicx,amssymb,amsmath,bm,latexsym}

\begin{document}


\vskip .5in

\begin{center}

{\Large\bf  Fifth-order generalized Heisenberg
supermagnetic models} \vskip .5in

{\large   Nana Jiang$^{a}$, Meina Zhang$^{b}$, Jiafeng Guo$^{c}$\footnote{Corresponding author: jiafeng\_guo@163.com}  and  Zhaowen Yan$^{a}$\footnote{Corresponding author: yanzw@imu.edu.cn}} \\
\vs{10}

$^a${\em School of Mathematical Sciences, Inner Mongolia University, Hohhot 010021, China}\\

$^b${\em School of Mathematics and Statistics, Northeast Normal University, Changchun 130024, China}\\

$^c${\em Department of Mathematical Sciences, Zibo Normal College, Zibo, 255130, China}\\
\end{center}
\vskip .2in \vspace{.3in}

\begin{abstract}

This paper is concerned with the construction of the fifth-order generalized Heisenberg supermagnetic models. We also investigate the integrable structure and properties of the supersymmetric systems. We establish their gauge equivalent equations with the gauge transformation for two quadratic constraints, i.e., the super fifth-order nonlinear Schr\"{o}dinger equation and the fermionic fifth-order nonlinear Schr\"{o}dinger equation, respectively.
 \end{abstract}

\section{Introduction}

The Heisenberg ferromagnet (HF) model \cite{LM77,ZVE}
 \begin{eqnarray}
 \mathbf{S}_t=\mathbf{S}\times \mathbf{S}_{xx},\ \ \mathbf{S}=(S_1,S_2,S_3),\ \ \mathbf{S}\cdot\mathbf{S}=1
 \end{eqnarray}
is the simplest integrable model of ferromagnetism, where $S$ is a spin vector and $\times$ means vector product. HF model attracts a great interest mainly due to its widely applications roles in various fields in mathematics and physics, for example, the anti-de Sitter/conformal field theories \cite{FG}-\cite{FGSLB}, the two-dimensional gravity theory \cite{LMK} and Eulerian vortex filament \cite{DB}.
The HF model is shown to be gauge and geometrical equivalent to the nonlinear Schr\"{o}dinger equation (NLSE) \cite{ZVE}. In terms of the inverse scattering method, Takhtajan \cite{Ta} discussed its Lax representation, the one-soliton solution and the phase and center of mass shifts for a two-soliton collision. A variety of the work has focused on the investigation of the generalized HF models, such as higher-order deformations of HF models \cite{AB,Va}, the multi-component
generalized HF models \cite{MA}, the multidimensional extended HF models \cite{MR1,MR}. Later on the N-soliton solutions of the generalized HF models have also been analyzed \cite{DB,FD}.

Supersymmetry has its origins in quantum field theory and it receives a lot of attention \cite{1}-\cite{5}. A variety of integrable systems have been generalized into their supersymmetric equations, for instance, the Korteweg-de Vries (KdV) equation \cite{Mathieu}, the Heisenberg supermagnet (HS) model \cite{Makhankov, Guo} and the NLSE  \cite{Yan1}. The Heisenberg supermagnetic (HS) model can be regarded as the super generalization of the HF model. Two integrable HS models are constructed on the compact subalgebra $su(2/1)$. Moreover, by virtue of gauge transformation, Makhankov and Pashaev \cite{Makhankov} proposed two types of gauge equivalent equations. It should be point out the HS model has connection with the strong electron correlated Hubbard model. Recently, a great deal of extended HS models have been studied \cite{Yan2,Yan3}. Furthermore, one \cite{Yan4,Yan5} established third-order and forth-order integrable HS models and presented their gauge equivalent counterparts. It is then a distinctive question to ask  what is the other higher-order HS models under two constraints. When we choose a Lax operator with a function of $S$, the difficulty of construction is how to derive another Lax operator with two quadratic constraints and contains fifth-order derivatives with respect to $x$. In this paper, we intend to construct fifth-order HS models and discuss the corresponding integrability in the light of the gauge transformation.

The present paper is built up as follows. A brief review of some elementary facts with the HS model are treated in Section 2. In Section 3, we establish the fifth-order HS model and derive their  equivalent equations with gauge transformation. Finally, the last Section will be devoted to a summary and discussion.

\section{Heisenberg Supermagnet Model}
The HS model can be read as
\cite{Makhankov}
\begin{equation}\label{HS}
iS_{t}=[S,S_{xx}],
\end{equation}
where $S$ is a superspin function which can be expressed as
\begin{eqnarray}
S&=&2\sum\limits_{a=1}^{4}S_{a}T_{a}+2\sum\limits_{a=5}^{8}C_{a}T_{a},\notag \\
&=&\left( {\begin{array}{*{20}c} S_3+S_4 & S_1-iS_2 & C_5-iC_6\\ S_1+iS_2 & -S_3+S_4 & C_7-iC_8 \\  C_5+iC_6& C_7+iC_8 & 2S_4 \\ \end{array}}\right),
\end{eqnarray}
where $S_{1},\ldots ,S_{4}$ and $C_{5},\ldots ,C_{8}$ are the bosonic and  fermionic variables, respectively. $ T_{1},\ldots , T_{4}$  and $T_{5},\ldots ,T_{8}$  are bosonic and fermionic generators of the superalgebra $su(2/1)$, respectively.

The concept of gauge equivalence yields connections between the completely integrable systems. The most useful advantage of such interconnections is that we understand the integrable properties of both two systems associated with gauge transformation. It is known that the  equivalent equations with gauge transformation can be embedded to their supersymmetric extensions. One proved that the HS models (\ref{HS}) are gauge equivalent to the flowing systems with two constraints \cite{Makhankov}

(I). $S^2=S$ for $S \in SU(2/1)/S(L(1/1)\times U(1))$
\begin{eqnarray}\label{SN}
&&i\Phi_t+\Phi_{xx}+2(\Phi\bar{\Phi}+\Psi\bar{\Psi})\Phi=0, \notag\\
&&i\Psi_t+\Psi_{xx}+2 \Phi\bar\Phi\Psi=0.
\end{eqnarray}

(II). $S^2=3S-2I$ for $S \in SU(2/1)/S(U(2)\times U(1))$
\begin{eqnarray}\label{odd}
&&i\Psi_{1t}+\Psi_{1xx}+2\bar\Psi_{2}\Psi_2\Psi_1=0, \notag\\
&&i\Psi_{2t}+\Psi_{2xx}+2\bar\Psi_{1}\Psi_1\Psi_2=0,
\end{eqnarray}
where $\Phi(x,t)$ is a bosonic variable and $\Psi,\Psi_1,\Psi_2$ are the fermionic ones.

\section{Fifth-order Heisenberg Supermagnet Model}

    Now we consider a fifth-order generalized HS model with constraint (I)
\begin{eqnarray} S_t=\varepsilon S_{xxxxx}+E(S_x,S_{xx},S_{xxx},S_{xxxx}),
\end{eqnarray}
where $\varepsilon$ is a parameter. The generalization function $E(S_x,S_{xx},S_{xxx},S_{xxxx})$
needs to be determined which satisfies the transformation equation
\begin{eqnarray}\label{C1}
SE+ES=E.
\end{eqnarray}
With the constraint (I), the superspin variable satisfies  $SS_tS=0$ and $S[S,S_{xx}]S=0$.

Based on the Lax representation of the forth-order generalized HS model \cite{Guo}, the Lax pair $G$ of the fifth-order HS model contains no higher than the fourth-order derivatives with respect to $x$. Let us introduce
\begin{eqnarray}\label{Lax1}
F&=&-i\lambda S,\notag\\
G&=&-i\lambda^{5}S+\lambda^{4}[S,S_x]+\sum_{i=1}^{n}\lambda^i T_i(S,S_x,S_{xx},S_{xxx},S_{xxxx}),
\end{eqnarray}
where $\lambda$ is a spectral parameter.

The Lax pair satisfies the zero-curvature equation
\begin{eqnarray} \label{zero1}
F_t-G_x+[F,G]=0.
\end{eqnarray}
Substituting (\ref{Lax1}) into (\ref{zero1}) and using the condition (\ref{C1}), we obtain
\begin{eqnarray}
E&=&-5(S_{xxxx}S_x+S_xS_{xxxx})-10(S_{xxx}S_{xx}+S_{xx}S_{xxx}-SS_{xxxx}S_x-SS_xS_{xxxx})\notag\\
&&+15(S_xS_xS_{xxx}+S_{xxx}S_xS_x)+20(S_{xx}S_xS_{xx}+S_xS_{xxx}S_x+SS_{xxx}S_{xx}+S\notag\\
&&S_{xx}S_{xxx})+25(S_xS_{xx}S_{xx}+S_{xx}S_{xx}S_x)+70(S_xS_xS_xS_xS_x+SS_{xx}S_xS_xS_x+\notag\\
&&SS_xS_{xx}S_xS_x+SS_xS_xS_{xx}S_x+SS_xS_xS_xS_{xx})-35(S_{xx}S_xS_xS_x+S_xS_{xx}S_xS_x\notag\\
&&+S_xS_xS_{xx}S_x+S_xS_xS_xS_{xx}),\notag\\
T_1&=&i(-S_{xxxx}+5S_{xxx}S_x+5S_{xx}S_{xx}+5S_xS_{xxx}-10SS_{xxx}S_x-10SS_{xx}S_{xx}-10\notag\\
&&SS_xS_{xxx}-15S_{xx}S_xS_x-10S_xS_{xx}S_x-5S_xS_xS_{xx}+35S_xS_xS_xS_x-70SS_x\notag\\
&&S_xS_xS_x),\notag\\
T_2&=&-[S,S_{xxx}]+[S_x,S_{xx}]+10S_xS_xS_x-20SS_xS_xS_x, \notag\\
T_3&=&i(S_{xx}-3S_xS_x+6SS_xS_x),\notag\\
T_4&=&T_5=\cdots=T_n=0.
\end{eqnarray}
From the Eq.(\ref{zero1}) and contrasting coefficients of the power of $\lambda$, we derive the higher-order HS model
\begin{eqnarray}\label{St2}
iS_t&=&i[S_{xxxxx}-5(S_{xxx}S_x+S_{xx}S_{xx}+S_xS_{xxx}-S_xS_xS_{xx})_x+10(SS_{xxx}S_x\notag\\
&&+SS_{xx}S_{xx}+SS_xS_{xxx}+S_xS_{xx}S_x)_x+15(S_{xx}S_xS_x)_x+35(2SS_xS_xS_xS_x\notag\\
&&-S_xS_xS_xS_x)_x].
\end{eqnarray}
The corresponding $F$ and $G$ are given by
\begin{eqnarray}\label{Lax2}
F &=&-i\lambda S,  \notag \\
G &=&-i\lambda^{5}S+\lambda^{4}[S,S_x]+i\lambda^{3}(S_{xx}-3S_xS_x+6SS_xS_x)+\lambda^{2}(-[S,S_{xxx}]+\notag\\
&&[S_x,S_{xx}]+10S_xS_xS_x-20SS_xS_xS_x)+i\lambda(-S_{xxxx}+5S_{xxx}S_x+5S_{xx}\notag\\
&&S_{xx}+5S_xS_{xxx}-10SS_{xxx}S_x-10SS_{xx}S_{xx}-10SS_xS_{xxx}-15S_{xx}S_xS_x\notag\\
&&-10S_xS_{xx}S_x-5S_xS_xS_{xx}+35S_xS_xS_xS_x-70SS_xS_xS_xS_x),
\end{eqnarray}
where $\lambda$ is a spectral parameter.

Now we turn to consider the gauge equivalent equation of (\ref{St2}), let us take
\begin{equation}\label{Sxt}
S(x,t)=g^{-1}\left( x,t\right) \Sigma g \left( x,t\right),
\end{equation}
where $g(x,t)\in SU(2/1)$.

Introducing the relation
\begin{equation}\label{J10}
J_{1}=g_{x}g^{-1},\ \  J_{0}=g_{t}g^{-1},
\end{equation}
 Eq.(\ref{J10}) satisfies
\begin{equation}\label{Jzero}
\partial _{t}J_{1}-\partial _{x}J_{0}+[J_{1},J_{0}]=0.
\end{equation}
The orthogonal direct sum decomposition of the super algebra $su(2/1)$ is as follows
\begin{equation}
\mathrm{L=L}^{\mathrm{(0)}}\oplus \mathrm{L}^{\mathrm{(1)}},  \label{eq:zj}
\end{equation}
where  $[\mathrm{L}^{\mathrm{(0)}}\mathrm{,L}^{\mathrm{(0)}
}]\subset \mathrm{L}^{\mathrm{(0)}}$, $[\mathrm{L}^{\mathrm{(0)}}\mathrm{,L}^{\mathrm{(1)}
}]\subset \mathrm{L}^{\mathrm{(1)}}$, $[\mathrm{L}^{\mathrm{(1)}}\mathrm{,L}^{\mathrm{(1)}
}]_+\subset \mathrm{L}^{\mathrm{(0)}}$. The commutation  and anticommutator relations are given by
 $[X,Y]=XY-YX$, $[X,Y]_-=XY+YX$. $L^{(0)}$ is an algebra constructed in terms of the stationary subgroup H.
Suppose
\begin{eqnarray}\label{J1}
 J_{1}&=&i\left( {\begin{array}{*{20}c} 0 & \varphi & \psi \\ {\bar \varphi }
& 0 & 0 \\ {\bar \psi } & 0 & 0 \\ \end{array}}\right)\in \mathrm{L}^{\mathrm{(1)}}\ \  \text{for} \ \ S \in SU(2/1)/S(L(1/1) \times U(1)),
\end{eqnarray}
where $\varphi (x,t)$ is bosonic filed and $\psi (x,t)$ is fermionic one.

Based on (\ref{Sxt}), (\ref{J10}) and (\ref{J1}), we have
\begin{eqnarray}\label{Sx}
S_{t} &=& g^{-1}(x,t)[\Sigma ,J_{0}]g(x,t), \notag\\
S_{x} &=&g^{-1}(x,t)[\Sigma ,J_{1}]g(x,t), \notag\\
S_{xx} &=&g^{-1}(x,t)([[\Sigma ,J_{1}],J_{1}]+[\Sigma ,J_{1x}])g(x,t),\notag\\
S_{xxx}&=&g^{-1}(x,t)([\Sigma,J_{1xx}]+[[[\Sigma,J_1],J_1],J_1]+2[[\Sigma,J_{1x}],J_1]+[[\Sigma,J_1],J_{1x}])\notag\\
&&g(x,t),\notag\\
S_{xxxx}&=&g^{-1}(x,t)(3[[[\Sigma,J_{1x}],J_1],J_1]+3[[\Sigma,J_{1xx}],J_1]+[[[[\Sigma,J_1],J_1],J_1],J_1])\notag\\
&&+2[[[\Sigma,J_1],J_{1x}],J_1]+3[[\Sigma,J_{1x}],J_{1x}]+[\Sigma,J_{1xxx}]+[[[\Sigma,J_1],J_1],J_{1x}]\notag\\
&&+[[\Sigma,J_1],J_{1xx}])g(x,t),\notag\\
S_{xxxxx}&=&g^{-1}(x,t)(4[[[[\Sigma,J_{1x}],J_1],J_1],J_1]+6[[[\Sigma,J_{1xx}],J_1],J_1]\notag\\
&&+[[[[[\Sigma,J_1],J_1],J_1],J_1],J_1]+3[[[[\Sigma,J_1],J_{1x},],J_1],J_1]+8[[[\Sigma,J_{1x}],J_{1x}],J_1]\notag\\
&&+4[[\Sigma,J_{1xxx}],J_1]+2[[[[\Sigma,J_1],J_1],J_{1x}],J_1]+3[[[\Sigma,J_1],J_{1xx}],J_1]\notag\\
&&+4[[[\Sigma,J_{1x}],J_1],J_{1x}]+6[[\Sigma,J_{1xx}],J_{1x}]+[[[[\Sigma,J_1],J_1],J_1],J_{1x}]\notag\\
&&+3[[[\Sigma,J_1],J_{1x}],J_{1x}]+4[[\Sigma,J_{1x}],J_{1xx}]+[\Sigma,J_{1xxxx}]+[[[\Sigma,J_1],J_1],J_{1xx}]\notag\\
&&+[[\Sigma,J_1],J_{1xxx}])g(x,t).
\end{eqnarray}
By substituting (\ref{Sxt}) and (\ref{Sx}) into (\ref{St2}), we have
\begin{eqnarray}\label{Sigma}
[\Sigma,J_0]&=&4[[[[\Sigma,J_{1x}],J_1],J_1],J_1]+6[[[\Sigma,J_{1xx}],J_1],J_1]+[[[[[\Sigma,J_1],J_1],J_1],J_1],J_1]\notag\\
&&+3[[[[\Sigma,J_1],J_{1x},],J_1],J_1]+8[[[\Sigma,J_{1x}],J_{1x}],J_1]+4[[\Sigma,J_{1xxx}],J_1]\notag\\
&&+2[[[[\Sigma,J_1],J_1],J_{1x}],J_1]+3[[[\Sigma,J_1],J_{1xx}],J_1]+4[[[\Sigma,J_{1x}],J_1],J_{1x}]\notag
\end{eqnarray}
\begin{eqnarray}
&&+6[[\Sigma,J_{1xx}],J_{1x}]+[[[[\Sigma,J_1],J_1],J_1],J_{1x}]+3[[[\Sigma,J_1],J_{1x}],J_{1x}]\notag\\
&&+4[[\Sigma,J_{1x}],J_{1xx}]+[\Sigma,J_{1xxxx}]+[[[\Sigma,J_1],J_1],J_{1xx}]+[[\Sigma,J_1],J_{1xxx}]\notag\\
&&-5(([\Sigma,J_{1xx}]+[[[\Sigma,J_1],J_1],J_1]+2[[\Sigma,J_{1x}],J_1]+[[\Sigma,J_1],J_{1x}])[\Sigma ,J_{1}]\notag\\
&&+([[\Sigma ,J_{1}],J_{1}]+[\Sigma ,J_{1x}])([[\Sigma ,J_{1}],J_{1}]+[\Sigma ,J_{1x}])+[\Sigma ,J_{1}]([\Sigma,J_{1xx}]\notag\\
&&+[[[\Sigma,J_1],J_1],J_1]+2[[\Sigma,J_{1x}],J_1]+[[\Sigma,J_1],J_{1x}])-[\Sigma ,J_{1}][\Sigma ,J_{1}]\notag\\
&&([[\Sigma ,J_{1}],J_{1}]+[\Sigma ,J_{1x}]))_x+10(\Sigma([\Sigma,J_{1xx}]+[[[\Sigma,J_1],J_1],J_1]\notag\\
&&+2[[\Sigma,J_{1x}],J_1]+[[\Sigma,J_1],J_{1x}])[\Sigma ,J_{1}]+\Sigma([[\Sigma ,J_{1}],J_{1}]+[\Sigma ,J_{1x}])\notag\\
&&([[\Sigma ,J_{1}],J_{1}]+[\Sigma ,J_{1x}])+\Sigma[\Sigma,J_{1}]([\Sigma,J_{1xx}]+[[[\Sigma,J_1],J_1],J_1]\notag\\ &&+2[[\Sigma,J_{1x}],J_1]+[[\Sigma,J_1],J_{1x}])+[\Sigma ,J_{1}]([[\Sigma ,J_{1}],J_{1}]+[\Sigma ,J_{1x}])[\Sigma ,J_{1}])_x\notag\\
&&+15(([[\Sigma ,J_{1}],J_{1}]+[\Sigma ,J_{1x}])[\Sigma ,J_{1}][\Sigma ,J_{1}])_x+35(2\Sigma[\Sigma ,J_{1}][\Sigma ,J_{1}][\Sigma ,J_{1}]\notag\\
&&[\Sigma ,J_{1}]-[\Sigma ,J_{1}][\Sigma ,J_{1}][\Sigma ,J_{1}][\Sigma ,J_{1}])_x,
\end{eqnarray}
where $\Sigma=diag(0,1,1)$.

From Eq.(\ref{Sigma}) and  $[\Sigma,J_0^{(0)}]=0$, we have

\begin{equation}\label{1J01}
J_{0}^{(1)}=\left(
\begin{array}{ccc}
0 & (J_{0}^{(1)})_{12} & (J_{0}^{(1)})_{13} \\
(J_{0}^{(1)})_{21} & 0 & 0 \\
(J_{0}^{(1)})_{31} & 0 & 0
\end{array}
\right),
\end{equation}
where
\begin{eqnarray}
(J_{0}^{(1)})_{12}&=&6i(\varphi\bar{\varphi}\varphi\bar{\varphi}\varphi+2\psi\bar{\psi}\varphi\bar{\varphi}\varphi)+i\varphi_{xxxx}+2i((\varphi\bar{\varphi}_x)_x\varphi+(\varphi_x\bar{\varphi})_x\varphi+\notag\\
&&3(\varphi_x\varphi)_x\bar{\varphi}+2(\psi_x\varphi)_x\bar{\psi}+(\psi\varphi_x)_x\bar{\psi}+\psi(\bar{\psi}\varphi_x)_x+(\psi\bar{\psi}_x)_x\varphi),\notag\\
(J_{0}^{(1)})_{13}&=&6i\varphi\bar{\varphi}\varphi\bar{\varphi}\psi+i\psi_{xxxx}+2i(2(\varphi_x\psi)_x\bar{\varphi}+(\varphi\bar{\varphi}_x)_x\psi+(\bar{\varphi}\psi_x)_x\varphi+\notag\\
&&(\varphi\psi_x)_x\bar{\varphi}),\notag\\
(J_{0}^{(1)})_{21}&=&6i(\bar{\varphi}\varphi\bar{\varphi}\varphi\bar{\varphi}+2\bar{\varphi}\varphi\bar{\varphi}\psi\bar{\psi})+i\bar{\varphi}_{xxxx}+2i((\varphi_x\bar{\varphi})_x\bar{\varphi}+(\varphi\bar{\varphi}_x)_x\bar{\varphi}+\notag\\
&&3(\bar{\varphi}_x\bar{\varphi})_x\varphi+2\psi(\bar{\varphi}\bar{\psi}_x)_x+\psi(\bar{\varphi}_x\bar{\psi})_x+(\bar{\varphi}_x\psi)_x\bar{\psi}+(\psi_x\bar{\psi})_x\bar{\varphi}),\notag\\
(J_{0}^{(1)})_{31}&=&6i\bar{\varphi}\varphi\bar{\varphi}\varphi\bar{\psi}+i\bar{\psi}_{xxxx}+2i(2(\bar{\varphi}_x\bar{\psi})_x\varphi+(\bar{\varphi}\varphi_x)_x\bar{\psi}+(\varphi\bar{\psi}_x)_x\bar{\varphi}+\notag\\
&&(\bar{\varphi}\bar{\psi}_x)_x\varphi).
\end{eqnarray}
According to the Eq.(\ref{Jzero}) and Eq.(\ref{eq:zj}), we obtain
\begin{eqnarray}\label{J0x}
(J_0^{(0)})_x=[J_1,J_0^{(1)}].
\end{eqnarray}
Substituting (\ref{J1}), (\ref{1J01}) into (\ref{J0x}) and integrating Eq.(\ref{J0x}) in reference to respect to the variable $x$, we derive

\begin{equation}
J_{0}^{(0)}=\left(
\begin{array}{ccc}
(J_{0}^{(0)})_{11} & 0 & 0 \\
0 &(J_{0}^{(0)})_{22} & (J_{0}^{(0)})_{23} \\
0 & (J_{0}^{(0)})_{32} & (J_{0}^{(0)})_{33}
\end{array}
\right) ,
\end{equation}
where
\begin{eqnarray}
(J_{0}^{(0)})_{11}&=&6(\varphi_x\varphi\bar{\varphi}\bar{\varphi}-\bar{\varphi}_x\bar{\varphi}\varphi\varphi+\varphi_x\psi\bar{\varphi}\bar{\psi}+\varphi\psi_x\bar{\varphi}\bar{\psi}-\psi\varphi\bar{\varphi}_x\bar{\psi}-\psi\varphi\bar{\varphi}\bar{\psi}_x)\notag\\
&&+\bar{\varphi}\varphi_{xxx}-\bar{\varphi}_{xxx}\varphi-\bar{\varphi}_x\varphi_{xx}+\bar{\varphi}_{xx}\varphi_x+\psi_{xxx}\bar{\psi}-\psi\bar{\psi}_{xxx}-\psi_{xx}\notag\\
&&\bar{\psi}_x+\psi_x\bar{\psi}_{xx},\notag\\
(J_{0}^{(0)})_{22}&=&6(\bar{\varphi}_x\bar{\varphi}\varphi\varphi-\varphi_x\varphi\bar{\varphi}\bar{\varphi})+4(\varphi\psi\bar{\varphi}_x\bar{\psi}-\varphi_x\psi\bar{\varphi}\bar{\psi})+2(\varphi\psi\bar{\varphi}\bar{\psi}_x-\psi_x\notag\\
&&\varphi\bar{\varphi}\bar{\psi})+\bar{\varphi}_{xxx}\varphi-\bar{\varphi}\varphi_{xxx}-\bar{\varphi}_{xx}\varphi_x+\bar{\varphi}_x\varphi_{xx},\notag\\
(J_{0}^{(0)})_{23}&=&4(\bar{\varphi}_x\bar{\varphi}\varphi\psi-\varphi\psi_x\bar{\varphi}\bar{\varphi})+2(\bar{\varphi}_x\bar{\varphi}\varphi\psi-\varphi_x\psi\bar{\varphi}\bar{\varphi}+\psi_x\bar{\psi}\bar{\varphi}\psi)+\bar{\varphi}_{xxx}\psi\notag\\
&&-\bar{\varphi}\psi_{xxx}-\bar{\varphi}_{xx}\psi_x+\bar{\varphi}_x\psi_{xx},\notag\\
(J_{0}^{(0)})_{32}&=&4(\bar{\varphi}\bar{\psi}_x\varphi\varphi-\varphi_x\varphi\bar{\varphi}\bar{\psi})+2(\bar{\varphi}_x\bar{\psi}\varphi\varphi-\varphi_x\varphi\bar{\varphi}\bar{\psi}-\bar{\psi}\varphi\psi\bar{\psi}_x)+\bar{\psi}_{xxx}\varphi\notag\\
&&-\bar{\psi}\varphi_{xxx}-\bar{\psi}_{xx}\varphi_x+\bar{\psi}_x\varphi_{xx},\notag\\
(J_{0}^{(0)})_{33}&=&4(\bar{\varphi}\bar{\psi}_x\varphi\psi-\bar{\psi}\bar{\varphi}\varphi\psi_x)+2(\bar{\varphi}_x\bar{\psi}\varphi\psi-\bar{\psi}\bar{\varphi}\varphi_x\psi)+\bar{\psi}_{xxx}\psi-\bar{\psi}\psi_{xxx}\notag\\
&&-\bar{\psi}_{xx}\psi_x+\bar{\psi}_x\psi_{xx}.
\end{eqnarray}
Since $J_0=J_0^{(0)}+J_0^{(1)}$, it is easy to draw the following conclusion
\begin{equation} \label{J0}
J_{0} =\left(
\begin{array}{ccc}
(J_{0}^{(0)})_{11}&(J_{0}^{(1)})_{12}  &(J_{0}^{(1)})_{13}\\
(J_{0}^{(1)})_{21} &(J_{0}^{(0)})_{22} & (J_{0}^{(0)})_{23}
\\
(J_{0}^{(1)})_{31}& (J_{0}^{(0)})_{32} & (J_{0}^{(0)})_{33}
\end{array}
\right).
\end{equation}
By virtue of the gauge transformation, $F$ and $G$ in (\ref{Lax2}) lead to $\hat{F}$ and $\hat{G}$, respectively.
\begin{eqnarray}\label{hatFG}
\hat{F}&=&gFg^{-1}+g_{x}g^{-1}=-i\lambda \Sigma +J_{1},\notag\\
\hat{G}&=&gGg^{-1}+g_{t}g^{-1}\notag\\
&=&g(-i\lambda^{5}S+\lambda^{4}[S,S_x]+i\lambda^{3}(S_{xx}-3S_xS_x+6SS_xS_x)+\lambda^{2}(-[S,S_{xxx}]\notag\\
&&+[S_x,S_{xx}]+10S_xS_xS_x-20SS_xS_xS_x)+i\lambda(-S_{xxxx}+5S_{xxx}S_x+5\notag\\
&&S_{xx}S_{xx}+5S_xS_{xxx}-10SS_{xxx}S_x-10SS_{xx}S_{xx}-10SS_xS_{xxx}-15S_{xx}\notag\\
&&S_xS_x-10S_xS_{xx}S_x-5S_xS_xS_{xx}+35S_xS_xS_xS_x-70SS_xS_xS_xS_x)g^{-1}\notag\\
&&+J_0.\
\end{eqnarray}
Substituting (\ref{J1}) and (\ref{J0}) into (\ref{hatFG}), we obtain

\begin{eqnarray}
\widehat{F} =i\left(
\begin{array}{ccc}
0 & \varphi & \psi \\
\bar{\varphi } & -\lambda & 0 \\
\bar{\psi } & 0 & -\lambda
\end{array}
\right) , \ \ \ \
\widehat{G} =\left(
\begin{array}{ccc}
\widehat{G}_{11} & \widehat{G}_{12} & \widehat{G}_{13} \\
\widehat{G}_{21} &\widehat{G}_{22} & \widehat{G}_{23} \\
\widehat{G}_{31} & \widehat{G}_{32} &\widehat{G}_{33}
\end{array}
\right) ,
\end{eqnarray}
where
\begin{eqnarray}
\widehat{G}_{11}&=&6(\varphi_x\varphi\bar{\varphi}\bar{\varphi}-\bar{\varphi}_x\bar{\varphi}\varphi\varphi+\varphi_x\psi\bar{\varphi}\bar{\psi}+\varphi\psi_x\bar{\varphi}\bar{\psi}-\psi\varphi\bar{\varphi}_x\bar{\psi}-\psi\varphi\bar{\varphi}\bar{\psi}_x)+\bar{\varphi}\varphi_{xxx}-\notag\\
&&\bar{\varphi}_{xxx}\varphi-\bar{\varphi}_x\varphi_{xx}+\bar{\varphi}_{xx}\varphi_x+\psi_{xxx}\bar{\psi}-\psi\bar{\psi}_{xxx}-\psi_{xx}\bar{\psi}_x+\psi_x\bar{\psi}_{xx}-i\lambda^{3}(\varphi\bar{\varphi}+\notag\\
&&\psi\bar{\psi})+\lambda^{2}(\varphi\bar{\varphi}_x+\psi\bar{\psi}_x-\varphi_x\bar{\varphi}-\psi_x\bar{\psi})+i\lambda(\varphi_{xx}\bar{\varphi}+\psi_{xx}\bar{\psi}+\varphi\bar{\varphi}_{xx}+\psi\bar{\psi}_{xx}-\notag\\
&&\varphi_x\bar{\varphi}_x-\psi_x\bar{\psi}_x)+3i\lambda\varphi\bar{\varphi}\varphi\bar{\varphi}+6i\lambda\varphi\bar{\varphi}\psi\bar{\psi},\notag\\
\widehat{G}_{12}&=&6i(\varphi\bar{\varphi}\varphi\bar{\varphi}\varphi+2\psi\bar{\psi}\varphi\bar{\varphi}\varphi)+i\varphi_{xxxx}+2i((\varphi\bar{\varphi}_x)_x\varphi+(\varphi_x\bar{\varphi})_x\varphi+3(\varphi_x\varphi)_x\bar{\varphi}+\notag\\
&&2(\psi_x\varphi)_x\bar{\psi}+(\psi\varphi_x)_x\bar{\psi}+\psi(\bar{\psi}\varphi_x)_x+(\psi\bar{\psi}_x)_x\varphi)+i\lambda^{4}\varphi+\lambda^{3}\varphi_x-i\lambda^{2}\varphi_{xx}-\notag\\
&&2i\lambda^{2}(\varphi\bar{\varphi}\varphi+\psi\bar{\psi}\varphi)-\lambda\varphi_{xxx}-6\lambda\varphi_x\bar{\varphi}\varphi-3\lambda\psi_x\bar{\psi}\varphi-3\lambda\psi\bar{\psi}\varphi_x,\notag\\
\widehat{G}_{13}&=&6i\varphi\bar{\varphi}\varphi\bar{\varphi}\psi+i\psi_{xxxx}+2i(2(\varphi_x\psi)_x\bar{\varphi}+(\varphi\bar{\varphi}_x)_x\psi+(\bar{\varphi}\psi_x)_x\varphi+(\varphi\psi_x)_x\bar{\varphi})+i\notag\\
&&\lambda^{4}\psi+\lambda^{3}\psi_x-i\lambda^{2}\psi_{xx}-2i\lambda^{2}\varphi\bar{\varphi}\psi-\lambda\psi_{xxx}-3\lambda\varphi_x\bar{\varphi}\psi-3\lambda\varphi\bar{\varphi}\psi_x,\notag\\
\widehat{G}_{21}&=&6i(\bar{\varphi}\varphi\bar{\varphi}\varphi\bar{\varphi}+2\bar{\varphi}\varphi\bar{\varphi}\psi\bar{\psi})+i\bar{\varphi}_{xxxx}+2i((\varphi_x\bar{\varphi})_x\bar{\varphi}+(\varphi\bar{\varphi}_x)_x\bar{\varphi}+3(\bar{\varphi}_x\bar{\varphi})_x\varphi+\notag\\
&&2\psi(\bar{\varphi}\bar{\psi}_x)_x+\psi(\bar{\varphi}_x\bar{\psi})_x+(\bar{\varphi}_x\psi)_x\bar{\psi}+(\psi_x\bar{\psi})_x\bar{\varphi})+i\lambda^{4}\bar{\varphi}-\lambda^{3}\bar{\varphi}_x-i\lambda^{2}\bar{\varphi}_{xx}-\notag\\
&&2i\lambda^{2}(\bar{\varphi}\varphi\bar{\varphi}+\bar{\varphi}\psi\bar{\psi})+\lambda\bar{\varphi}_{xxx}+6\lambda\bar{\varphi}_x\varphi\bar{\varphi}+3\lambda\bar{\varphi}_x\psi\bar{\psi}+3\lambda\bar{\varphi}\psi\bar{\psi}_x,\notag\\
\widehat{G}_{22}&=&6(\bar{\varphi}_x\bar{\varphi}\varphi\varphi-\varphi_x\varphi\bar{\varphi}\bar{\varphi})+4(\varphi\psi\bar{\varphi}_x\bar{\psi}-\varphi_x\psi\bar{\varphi}\bar{\psi})+2(\varphi\psi\bar{\varphi}\bar{\psi}_x-\psi_x\varphi\bar{\varphi}\bar{\psi})+\bar{\varphi}_{xxx}\varphi\notag\\
&&-\bar{\varphi}\varphi_{xxx}-\bar{\varphi}_{xx}\varphi_x+\bar{\varphi}_x\varphi_{xx}-i\lambda^{5}+i\lambda^{3}\varphi\bar{\varphi}+\lambda^{2}(\bar{\varphi}\varphi_x-\bar{\varphi}_x\varphi)-i\lambda(\bar{\varphi}_{xx}\varphi+\bar{\varphi}\notag\\
&&\varphi_{xx}-\bar{\varphi}_x\varphi_x)+5i\lambda(\bar{\varphi}\varphi\bar{\varphi}\varphi+\bar{\varphi}\psi\bar{\psi}\varphi)-8i\lambda\bar{\varphi}\varphi\bar{\varphi}\varphi-8i\lambda\bar{\varphi}\psi\bar{\psi}\varphi,\notag\\
\widehat{G}_{23}&=&4(\bar{\varphi}_x\bar{\varphi}\varphi\psi-\varphi\psi_x\bar{\varphi}\bar{\varphi})+2(\bar{\varphi}_x\bar{\varphi}\varphi\psi-\varphi_x\psi\bar{\varphi}\bar{\varphi}+\psi_x\bar{\psi}\bar{\varphi}\psi)+\bar{\varphi}_{xxx}\psi-\bar{\varphi}\psi_{xxx}-\notag\\
&&\bar{\varphi}_{xx}\psi_x+\bar{\varphi}_x\psi_{xx}+i\lambda^{3}\bar{\varphi}\psi+\lambda^{2}(\bar{\varphi}\psi_x-\bar{\varphi}_x\psi)-i\lambda(\bar{\varphi}_{xx}\psi+\bar{\varphi}\psi_{xx}-\bar{\varphi}_x\psi_x)-\notag\\
&&3i\lambda\bar{\varphi}\varphi\bar{\varphi}\psi,\notag\\
\widehat{G}_{31}&=&6i\bar{\varphi}\varphi\bar{\varphi}\varphi\bar{\psi}+i\bar{\psi}_{xxxx}+2i(2(\bar{\varphi}_x\bar{\psi})_x\varphi+(\bar{\varphi}\varphi_x)_x\bar{\psi}+(\varphi\bar{\psi}_x)_x\bar{\varphi}+(\bar{\varphi}\bar{\psi}_x)_x\varphi)+i\lambda^{4}\notag\\
&&\bar{\psi}-\lambda^{3}\bar{\psi}_x-i\lambda^{2}\bar{\psi}_{xx}-2i\lambda^{2}\bar{\psi}\varphi\bar{\varphi}+\lambda\bar{\psi}_{xxx}+3\lambda\bar{\psi}_x\varphi\bar{\varphi}+3\lambda\bar{\psi}\varphi\bar{\varphi}_x,\notag\\
\widehat{G}_{32}&=&4(\bar{\varphi}\bar{\psi}_x\varphi\varphi-\varphi_x\varphi\bar{\varphi}\bar{\psi})+2(\bar{\varphi}_x\bar{\psi}\varphi\varphi-\varphi_x\varphi\bar{\varphi}\bar{\psi}-\bar{\psi}\varphi\psi\bar{\psi}_x)+\bar{\psi}_{xxx}\varphi-\bar{\psi}\varphi_{xxx}-\notag\\
&&\bar{\psi}_{xx}\varphi_x+\bar{\psi}_x\varphi_{xx}+i\lambda^{3}\bar{\psi}\varphi+\lambda^{2}(\bar{\psi}\varphi_x-\bar{\psi}_x\varphi)-i\lambda(\bar{\psi}\varphi_{xx}+\bar{\psi}_{xx}\varphi-\bar{\psi}_x\varphi_x)-\notag\\
&&3i\lambda\bar{\psi}\varphi\bar{\varphi}\varphi,\notag\\
\widehat{G}_{33}&=&4(\bar{\varphi}\bar{\psi}_x\varphi\psi-\bar{\psi}\bar{\varphi}\varphi\psi_x)+2(\bar{\varphi}_x\bar{\psi}\varphi\psi-\bar{\psi}\bar{\varphi}\varphi_x\psi)+\bar{\psi}_{xxx}\psi-\bar{\psi}\psi_{xxx}-\bar{\psi}_{xx}\psi_x+\notag\\
&&\bar{\psi}_x\psi_{xx}-i\lambda^{5}+i\lambda^{3}\bar{\psi}\psi+\lambda^{2}(\bar{\psi}\psi_x-\bar{\psi}_x\psi)-i\lambda(\bar{\psi}\psi_{xx}+\bar{\psi}_{xx}\psi-\bar{\psi}_x\psi)-3i\lambda\notag\\
&&\bar{\psi}\varphi\bar{\varphi}\psi.
\end{eqnarray}
By means of the zero-curvature formulation of $\hat{F}$ and $\hat{G}$, we derive the super fifth-order NLSE with the constraint (I)
\begin{eqnarray}
&&i\varphi_{t}-6i(\varphi\bar{\varphi}\varphi\bar{\varphi}\varphi+2\psi\bar{\psi}\varphi\bar{\varphi}\varphi)_x-i\varphi_{xxxxx}-2i((\varphi\bar{\varphi}_x)_x\varphi+(\varphi_x\bar{\varphi})_x\varphi+3(\varphi_x\varphi)_x\bar{\varphi}+\notag
\end{eqnarray}
\begin{eqnarray}
&&2(\psi_x\varphi)_x\bar{\psi}+(\psi\varphi_x)_x\bar{\psi}+\psi(\bar{\psi}\varphi_x)_x+(\psi\bar{\psi}_x)_x\varphi)_x+12i(\varphi\bar{\varphi}_x\varphi\bar{\varphi}\varphi-\varphi_x\varphi\varphi\bar{\varphi}\varphi)-16i\notag\\
&&\varphi_x\bar{\varphi}\varphi\psi\bar{\psi}+12i(\varphi\bar{\varphi}_x\varphi\psi\bar{\psi}+\varphi\varphi\bar{\varphi}\psi\bar{\psi}_x)-8i\varphi\varphi\bar{\varphi}\psi_x\bar{\psi}+i(2\varphi\bar{\varphi}_{xxx}\varphi-2\varphi\bar{\varphi}\varphi_{xxx}-2\varphi\notag\\
&&\bar{\varphi}_{xx}\varphi_x+2\varphi\bar{\varphi}_x\varphi_{xx}+2\psi\bar{\psi}_{xxx}\varphi-\psi\bar{\psi}\varphi_{xxx}-\psi\bar{\psi}_{xx}\varphi_x+\psi\bar{\psi}_x\varphi_{xx}-\psi_{xxx}\bar{\psi}\varphi+\psi_{xx}\bar{\psi}_x\notag\\
&&\varphi-\psi_x\bar{\psi}_{xx}\varphi)=0,\notag
\end{eqnarray}
\begin{eqnarray}\label{NLSE}
&&i\psi_{t}-6i(\varphi\bar{\varphi}\varphi\bar{\varphi}\psi)_x-i\psi_{xxxxx}-2i(2(\varphi_x\psi)_x\bar{\varphi}+(\varphi\bar{\varphi}_x)_x\psi+(\bar{\varphi}\psi_x)_x\varphi+(\varphi\psi_x)_x\bar{\varphi})_x\notag\\
&&+12i\varphi\bar{\varphi}\varphi\bar{\varphi}_x\psi-8i\varphi_x\bar{\varphi}\varphi\bar{\varphi}\psi-4i\varphi\bar{\varphi}\varphi\bar{\varphi}\psi_x+i(2\varphi\bar{\varphi}_{xxx}\psi-\varphi\bar{\varphi}\psi_{xxx}-\varphi\bar{\varphi}_{xx}\psi_x+\varphi\notag\\
&&\bar{\varphi}_x\psi_{xx}-2\psi\bar{\psi}_{xx}\psi_x+2\psi\bar{\psi}_x\psi_{xx}-\varphi_{xxx}\bar{\varphi}\psi-\varphi_x\bar{\varphi}_{xx}\psi+\varphi_{xx}\bar{\varphi}_x\psi)=0.
\end{eqnarray}

In the following, we turn to consider the second constraint  $S^2=3S-2I$. One derives $S_t$ and $[S, S_{xx}]$ satisfying $SS_tS=2S_t$ and $S[S, S_{xx}]S =2[S, S_{xx}]$. Thus the deformation term $E$ should satisfy the equation
\begin{eqnarray}\label{C2}
SE+ES=3E.
\end{eqnarray}

Using the similar approach as the above constraint, we obtain the fifth-order HS model under the constraint (II)
\begin{eqnarray}\label{St3}
S_t&=&S_{xxxxx}-15(S_{xxx}S_x+S_{xx}S_{xx}+S_xS_{xxx}-S_{xx}S_xS_x)_x+10(SS_{xxx}S_x\notag\\
&&+SS_{xx}S_{xx}+SS_xS_{xxx}+S_xS_{xx}S_x)_x+5(S_xS_xS_{xx})_x-105(S_xS_xS_xS_x)_x\notag\\
&&+70(SS_xS_xS_xS_x)_x.
\end{eqnarray}
The corresponding $F$ and $G$ can be expressed as
\begin{eqnarray}\label{FG3}
F &=&-i\lambda S,  \notag\\
G &=&-i\lambda^{5}S+\lambda^{4}[S,S_x]+i\lambda^{3}(S_{xx}-9S_xS_x+6SS_xS_x)+\lambda^{2}(-[S,S_{xxx}]+[S_x,S_{xx}]\notag\\
&&+30S_xS_xS_x-20SS_xS_xS_x)+i\lambda(-S_{xxxx}+15S_{xxx}S_x+15S_{xx}S_{xx}+15S_xS_{xxx}\notag\\
&&-10SS_{xxx}S_x-10SS_{xx}S_{xx}-10SS_xS_{xxx}-15S_{xx}S_xS_x-10S_xS_{xx}S_x-5S_xS_x\notag\\
&&S_{xx}+105S_xS_xS_xS_x-70SS_xS_xS_xS_x),
\end{eqnarray}
where $\lambda$ is the spectral parameter.

Let
\begin{eqnarray}\label{2J0}
 J_{1}&=&i\left( {\begin{array}{*{20}c} 0 & 0 & \psi_1 \\ {0 }
& 0 & \psi_2 \\ {\bar \psi_1 } & {\bar \psi_2 } & 0 \\ \end{array}}\right)\ \in \mathrm{L}^{\mathrm{(1)}} \  \text{for} \ \  S \in SU(2/1)/S(U(2) \times U(1)),
\end{eqnarray}
here $\psi_1(x,t),\psi_2(x,t)$ are the fermionic variables.

Substituting (\ref{Sxt}) and (\ref{Sx}) into (\ref{St3}), we find
\begin{eqnarray}\label{Sigma2}
[\Sigma,J_0]&=&4[[[[\Sigma,J_{1x}],J_1],J_1],J_1]+6[[[\Sigma,J_{1xx}],J_1],J_1]+[[[[[\Sigma,J_1],J_1],J_1],J_1],J_1]\notag\\
&&+3[[[[\Sigma,J_1],J_{1x},],J_1],J_1]+8[[[\Sigma,J_{1x}],J_{1x}],J_1]+4[[\Sigma,J_{1xxx}],J_1]\notag\\
&&+2[[[[\Sigma,J_1],J_1],J_{1x}],J_1]+3[[[\Sigma,J_1],J_{1xx}],J_1]+4[[[\Sigma,J_{1x}],J_1],J_{1x}]\notag\\
&&+6[[\Sigma,J_{1xx}],J_{1x}]+[[[[\Sigma,J_1],J_1],J_1],J_{1x}]+3[[[\Sigma,J_1],J_{1x}],J_{1x}]\notag\\
&&+4[[\Sigma,J_{1x}],J_{1xx}]+[\Sigma,J_{1xxxx}]+[[[\Sigma,J_1],J_1],J_{1xx}]+[[\Sigma,J_1],J_{1xxx}]\notag\\
&&-15(([\Sigma,J_{1xx}]+[[[\Sigma,J_1],J_1],J_1]+2[[\Sigma,J_{1x}],J_1]+[[\Sigma,J_1],J_{1x}])[\Sigma ,J_{1}]\notag\\
&&+([[\Sigma ,J_{1}],J_{1}]+[\Sigma ,J_{1x}])([[\Sigma ,J_{1}],J_{1}]+[\Sigma ,J_{1x}])+[\Sigma ,J_{1}]([\Sigma,J_{1xx}]\notag\\
&&+[[[\Sigma,J_1],J_1],J_1]+2[[\Sigma,J_{1x}],J_1]+[[\Sigma,J_1],J_{1x}])-([[\Sigma ,J_{1}],J_{1}]+[\Sigma ,J_{1x}])\notag\\
&&[\Sigma ,J_{1}][\Sigma ,J_{1}])_x+10(\Sigma([\Sigma,J_{1xx}]+[[[\Sigma,J_1],J_1],J_1]+2[[\Sigma,J_{1x}],J_1]\notag\\
&&+[[\Sigma,J_1],J_{1x}])[\Sigma ,J_{1}]+\Sigma([[\Sigma ,J_{1}],J_{1}]+[\Sigma ,J_{1x}])([[\Sigma ,J_{1}],J_{1}]+[\Sigma ,J_{1x}])\notag\\
&&+\Sigma[\Sigma,J_{1}]([\Sigma,J_{1xx}]+[[[\Sigma,J_1],J_1],J_1]+2[[\Sigma,J_{1x}],J_1]+[[\Sigma,J_1],J_{1x}])\notag\\
&&+[\Sigma ,J_{1}]([[\Sigma ,J_{1}],J_{1}]+[\Sigma ,J_{1x}])[\Sigma ,J_{1}])_x+5([\Sigma ,J_{1}][\Sigma ,J_{1}]([[\Sigma ,J_{1}],J_{1}]\notag\\
&&+[\Sigma ,J_{1x}]))_x-105([\Sigma ,J_{1}][\Sigma ,J_{1}][\Sigma ,J_{1}][\Sigma ,J_{1}])_x+70(\Sigma[\Sigma ,J_{1}][\Sigma ,J_{1}]\notag\\
&&[\Sigma ,J_{1}][\Sigma ,J_{1}])_x,
\end{eqnarray}
where $\Sigma=diag(1,1,2)$.\\
Repeating the process of constraint I, naturally, we obtain
\begin{eqnarray}\label{J01}
J _0^{(0)}&=&\left(
\begin{array}{ccc}
(J _0^{(0)})_{11}&(J _0^{(0)})_{12}& 0  \\
(J _0^{(0)})_{21}&(J _0^{(0)})_{22} & 0 \\
0 & 0 & (J _0^{(0)})_{33}
\end{array}\right),\ \
J _0^{(1)}=\left(
\begin{array}{ccc}
0 & 0 & (J _0^{(1)})_{13}\\
0 & 0 & (J _0^{(1)})_{23}\\
(J _0^{(1)})_{31}&(J_0^{(1)})_{32} & 0
\end{array}
\right).
\end{eqnarray}
Combining the two matrix in (\ref{J01}), we obtain
\begin{eqnarray}\label{2J1}
J _0&=&\left(
\begin{array}{ccc}
(J _0^{(0)})_{11}&(J _0^{(0)})_{12}&(J_0^{(1)})_{13}\\
(J _0^{(0)})_{21}&(J _0^{(0)})_{22}&(J_0^{(1)})_{23}\\
(J_0^{(1)})_{31}& (J_0^{(1)})_{32} &(J _0^{(0)})_{33}
\end{array}
\right),
\end{eqnarray}
where
\begin{eqnarray}
(J_0^{(0)})_{11}&=&4(\psi_1\psi_2\bar{\psi}_2\bar{\psi}_{1x}-\psi_{1x}\psi_2\bar{\psi}_2\bar{\psi}_1)-2(\bar{\psi}_2\bar{\psi}_1\psi_1\psi_{2x}-\psi_1\psi_2\bar{\psi}_{2x}\bar{\psi}_1)+\notag\\
&&\psi_{1xxx}\bar{\psi}_1-\psi_1\bar{\psi}_{1xxx}-\psi_{1xx}\bar{\psi}_{1x}+\psi_{1x}\bar{\psi}_{1xx},\notag\\
(J_0^{(0)})_{12}&=&2(\psi_1\bar{\psi}_{2x}\psi_2\bar{\psi}_2-\psi_1\bar{\psi}_1\psi_{1x}\bar{\psi}_2)+\psi_{1xxx}\bar{\psi}_2-\psi_1\bar{\psi}_{2xxx}-\psi_{1xx}\bar{\psi}_{2x}\notag\\
&&+\psi_{1x}\bar{\psi}_{2xx},\notag\\
(J_0^{(1)})_{13}&=&-4i(\psi_{1x}\psi_2)_x\bar{\psi}_2-2i(\bar{\psi}_2(\psi_1\psi_{2x})_x+\psi_1(\psi_{2x}\bar{\psi}_2)_x-(\psi_1\bar{\psi}_{2x})_x\psi_2)\notag\\
&&+i\psi_{1xxxx},\notag
\end{eqnarray}
\begin{eqnarray}
(J_0^{(0)})_{21}&=&2(\psi_2\bar{\psi}_{1x}\psi_1\bar{\psi}_1-\psi_2\bar{\psi}_2\psi_{2x}\bar{\psi}_1)+\psi_{2xxx}\bar{\psi}_1-\psi_2\bar{\psi}_{1xxx}-\psi_{2xx}\bar{\psi}_{1x}\notag\\
&&+\psi_{2x}\bar{\psi}_{1xx},\notag\\
(J_0^{(0)})_{22}&=&4(\psi_2\bar{\psi}_1\bar{\psi}_{2x}\psi_1-\psi_{2x}\psi_1\bar{\psi}_1\bar{\psi}_2)-2(\bar{\psi}_1\psi_2\psi_{1x}\bar{\psi}_2-\psi_2\bar{\psi}_{1x}\bar{\psi}_2\psi_1)+\notag\\
&&\psi_{2xxx}\bar{\psi}_2-\psi_2\bar{\psi}_{2xxx}-\psi_{2xx}\bar{\psi}_{2x}+\psi_{2x}\bar{\psi}_{2xx},\notag\\
(J_0^{(1)})_{23}&=&-4i(\psi_{2x}\psi_1)_x\bar{\psi}_1-2i(\bar{\psi}_1(\psi_2\psi_{1x})_x-\psi_2(\bar{\psi}_1\psi_{1x})_x-(\psi_2\bar{\psi}_{1x})_x\psi_1)\notag\\
&&+i\psi_{2xxxx},\notag\\
(J_0^{(1)})_{31}&=&-4i(\bar{\psi}_2\bar{\psi}_{1x})_x\psi_2+2i(\bar{\psi}_1(\bar{\psi}_{2x}\psi_2)_x+\bar{\psi}_2(\psi_{2x}\bar{\psi}_1)_x-(\bar{\psi}_{2x}\bar{\psi}_1)_x\psi_2)\notag\\
&&+i\bar{\psi}_{1xxxx},\notag\\
(J_0^{(1)})_{32}&=&-4i(\bar{\psi}_1\bar{\psi}_{2x})_x\psi_1+2i(\bar{\psi}_2(\bar{\psi}_{1x}\psi_1)_x+\bar{\psi}_1(\psi_{1x}\bar{\psi}_2)_x-(\bar{\psi}_{1x}\bar{\psi}_2)_x\psi_1)\notag\\
&&+i\bar{\psi}_{2xxxx},\notag\\
(J_{0}^{(0)})_{33}&=&6(\bar{\psi}_1\psi_{1x}\psi_2\bar{\psi}_2+\bar{\psi}_2\psi_{2x}\psi_1\bar{\psi}_1-\bar{\psi}_2\bar{\psi}_{1x}\psi_2\psi_1-\bar{\psi}_1\bar{\psi}_{2x}\psi_1\psi_2)+\bar{\psi}_{1xxx}\notag\\
&&\psi_1-\bar{\psi}_1\psi_{1xxx}-\bar{\psi}_{1xx}\psi_{1x}+\bar{\psi}_{1x}\psi_{1xx}+\bar{\psi}_{2xxx}\psi_2-\bar{\psi}_2\psi_{2xxx}-\bar{\psi}_{2xx}\notag\\
&&\psi_{2x}+\bar{\psi}_{2x}\psi_{2xx}.
\end{eqnarray}
By virtue of the gauge transformation, $F$ and $G$ (\ref{FG3}) lead to $\tilde{F}$ and $\tilde{G}$
\begin{eqnarray}\label{tildeFG}
\tilde{F}&=&gUg^{-1}+g_{x}g^{-1}=-i\lambda \Sigma +J_{1},  \notag \\
\tilde{G}&=&gVg^{-1}+g_{t}g^{-1}\notag\\
&=&g(-i\lambda^{5}S+\lambda^{4}[S,S_x]+i\lambda^{3}(S_{xx}-3S_xS_x+6SS_xS_x)+\lambda^{2}(-[S,S_{xxx}]\notag\\
&&+[S_x,S_{xx}]+10S_xS_xS_x-20SS_xS_xS_x)+i\lambda(-S_{xxxx}+5S_{xxx}S_x+5\notag\\
&&S_{xx}S_{xx}+5S_xS_{xxx}-10SS_{xxx}S_x-10SS_{xx}S_{xx}-10SS_xS_{xxx}-15S_{xx}\notag\\
&&S_xS_x-10S_xS_{xx}S_x-5S_xS_xS_{xx}+35S_xS_xS_xS_x-70SS_xS_xS_xS_x))g^{-1}\notag\\
&&+J_0.
\end{eqnarray}
By means of (\ref{2J0}) and (\ref{2J1}), we rewrite (\ref{tildeFG}) as follows
\begin{eqnarray}
\tilde{F}=i\left(
\begin{array}{ccc}
-\lambda & 0 & \psi_1 \\
0 & -\lambda & \psi_2 \notag\\
\bar{\psi}_1 & \bar{\psi_2} & -2\lambda
\end{array}
\right) , \ \ \
\tilde{G}=\left(
\begin{array}{ccc}
\tilde{G}_{11} & \tilde{G}_{12}& \tilde{G}_{13} \\
\tilde{G}_{21}&\tilde{G}_{22} &\tilde{G}_{23} \\
\tilde{G}_{31} & \tilde{G}_{32} &\tilde{G}_{33}
\end{array}
\right) ,\\
\end{eqnarray}
where
\begin{eqnarray}
\tilde{G}_{11}&=&-i\lambda^{5}-i\lambda^3\psi_1\bar{\psi}_1-\lambda^{2}(\psi_{1x}\bar{\psi}_1-\psi_1\bar{\psi}_{1x})+i\lambda(\psi_1\bar{\psi}_{1xx}+\psi_{1xx}\bar{\psi}_1)\notag\\
&&-i\lambda\psi_{1x}\bar{\psi}_{1x}+3i\lambda\psi_1\bar{\psi}_2\psi_2\bar{\psi}_1+4(\psi_1\psi_2\bar{\psi}_2\bar{\psi}_{1x}-\psi_{1x}\psi_2\bar{\psi}_2\bar{\psi}_1)-2\notag\\
&&(\bar{\psi}_2\bar{\psi}_1\psi_1\psi_{2x}-\psi_1\psi_2\bar{\psi}_{2x}\bar{\psi}_1)+(\psi_{1xxx}\bar{\psi}_1-\psi_1\bar{\psi}_{1xxx}-\psi_{1xx}\bar{\psi}_{1x}+\notag\\
&&\psi_{1x}\bar{\psi}_{1xx}),\notag
\end{eqnarray}
\begin{eqnarray}
\tilde{G}_{12}&=&-i\lambda^{3}\psi_1\bar{\psi}_2-\lambda^{2}(\psi_{1x}\bar{\psi}_2-\psi_1\bar{\psi}_{2x})+i\lambda(\psi_{1xx}\bar{\psi}_2+\psi_1\bar{\psi}_{2xx})-i\lambda\psi_{1x}\notag\\
&&\bar{\psi}_{2x}+2(\psi_1\bar{\psi}_{2x}\psi_2\bar{\psi}_2-\psi_1\bar{\psi}_1\psi_{1x}\bar{\psi}_2)+(\psi_{1xxx}\bar{\psi}_2-\psi_1\bar{\psi}_{2xxx}-\psi_{1xx}\notag\\
&&\bar{\psi}_{2x}+\psi_{1x}\bar{\psi}_{2xx}),\notag\\
\tilde{G}_{13}&=&i\lambda^{4}\psi_1+\lambda^{3}\psi_{1x}-i\lambda^{2}\psi_{1xx}+2i\lambda^{2}\psi_1\psi_2\bar{\psi}_2-3\lambda(\psi_{1x}\bar{\psi}_2\psi_2+\psi_1\bar{\psi}_2\psi_{2x})\notag\\
&&-\lambda\psi_{1xxx}-4i(\psi_{1x}\psi_2)_x\bar{\psi}_2-2i(\bar{\psi}_2(\psi_1\psi_{2x})_x+\psi_1(\psi_{2x}\bar{\psi}_2)_x-(\psi_1\bar{\psi}_{2x}\notag\\
&&)_x\psi_2)+i\psi_{1xxxx},\notag\\
\tilde{G}_{21}&=&-i\lambda^{3}\psi_2\bar{\psi}_1-\lambda^{2}(\psi_{2x}\bar{\psi}_1-\psi_2\bar{\psi}_{1x})+i\lambda(\psi_{2xx}\bar{\psi}_1+\psi_2\bar{\psi}_{1xx}-\psi_{2x}\bar{\psi}_{1x})\notag\\
&&+2(\psi_2\bar{\psi}_{1x}\psi_1\bar{\psi}_1-\psi_2\bar{\psi}_2\psi_{2x}\bar{\psi}_1)+(\psi_{2xxx}\bar{\psi}_1-\psi_2\bar{\psi}_{1xxx}-\psi_{2xx}\bar{\psi}_{1x}\notag\\
&&+\psi_{2x}\bar{\psi}_{1xx}),\notag\\
\tilde{G}_{22}&=&-i\lambda^{5}-i\lambda^{3}\psi_2\bar{\psi}_2-\lambda^{2}(\psi_{2x}\bar{\psi}_2-\psi_2\bar{\psi}_{2x})+i\lambda(\psi_2\bar{\psi}_{2xx}+\psi_{2xx}\bar{\psi}_2-\psi_{2x}\notag\\
&&\bar{\psi}_{2x})+3i\lambda\psi_2\bar{\psi}_1\psi_1\bar{\psi}_2+4(\psi_2\bar{\psi}_1\bar{\psi}_{2x}\psi_1-\psi_{2x}\psi_1\bar{\psi}_1\bar{\psi}_2)-2(\bar{\psi}_1\psi_2\psi_{1x}\bar{\psi}_2\notag\\
&&-\psi_2\bar{\psi}_{1x}\bar{\psi}_2\psi_1)+(\psi_{2xxx}\bar{\psi}_2-\psi_2\bar{\psi}_{2xxx}-\psi_{2xx}\bar{\psi}_{2x}+\psi_{2x}\bar{\psi}_{2xx}),\notag\\
\tilde{G}_{23}&=&i\lambda^{4}\psi_2+\lambda^{3}\psi_{2x}-i\lambda^{2}\psi_{2xx}-2i\lambda^{2}\psi_2\bar{\psi}_1\psi_1-3\lambda(\psi_{2}\bar{\psi}_1\psi_{1x}+\psi_{2x}\bar{\psi}_1\psi_{1})\notag\\
&&-\lambda\psi_{2xxx}-4i(\psi_{2x}\psi_1)_x\bar{\psi}_1-2i(\bar{\psi}_1(\psi_2\psi_{1x})_x-\psi_2(\bar{\psi}_1\psi_{1x})_x-(\psi_2\bar{\psi}_{1x})_x\notag\\
&&\psi_1)+i\psi_{2xxxx},\notag\\
\tilde{G}_{31}&=&i\lambda^{4}\bar{\psi}_1-\lambda^{3}\bar{\psi}_{1x}-i\lambda^{2}\bar{\psi}_{1xx}-2i\lambda^{2}\bar{\psi}_2\psi_2\bar{\psi}_1+3\lambda(\bar{\psi}_{2x}\psi_2\bar{\psi}_1+\bar{\psi}_2\psi_2\bar{\psi}_{1x})\notag\\
&&+\lambda\bar{\psi}_{1xxx}-4i(\bar{\psi}_2\bar{\psi}_{1x})_x\psi_2+2i(\bar{\psi}_1(\bar{\psi}_{2x}\psi_2)_x+\bar{\psi}_2(\psi_{2x}\bar{\psi}_1)_x-(\bar{\psi}_{2x}\bar{\psi}_1)_x\notag\\
&&\psi_2)+i\bar{\psi}_{1xxxx},\notag\\
\tilde{G}_{32}&=&i\lambda^{4}\bar{\psi}_2-\lambda^{3}\bar{\psi}_{2x}-i\lambda^{2}\bar{\psi}_{2xx}-2i\lambda^{2}\bar{\psi}_1\psi_1\bar{\psi}_2+3\lambda(\bar{\psi}_{1x}\psi_1\bar{\psi}_2+\bar{\psi}_1\psi_1\bar{\psi}_{2x})\notag\\
&&+\lambda\bar{\psi}_{2xxx}-4i(\bar{\psi}_1\bar{\psi}_{2x})_x\psi_1+2i(\bar{\psi}_2(\bar{\psi}_{1x}\psi_1)_x+\bar{\psi}_1(\psi_{1x}\bar{\psi}_2)_x-(\bar{\psi}_{1x}\bar{\psi}_2)_x\notag\\
&&\psi_1)+i\bar{\psi}_{2xxxx},\notag\\
\tilde{G}_{33}&=&-2i\lambda^{5}+i\lambda^{3}(\bar{\psi}_1\psi_1+\bar{\psi}_2\psi_2)-\lambda^{2}(\bar{\psi}_{1x}\psi_1-\bar{\psi}_1\psi_{1x}+\bar{\psi}_{2x}\psi_2-\bar{\psi}_2\psi_{2x})\notag\\
&&-i\lambda(\bar{\psi}_{1xx}\psi_1+\bar{\psi}_{2xx}\psi_2+\bar{\psi}_2\psi_{2xx}+\bar{\psi}_1\psi_{1xx}-\bar{\psi}_{1x}\psi_{1x}-\bar{\psi}_{2x}\psi_{2x})-6\notag\\
&&i\lambda\bar{\psi}_1\psi_1\bar{\psi}_2\psi_2+6(\bar{\psi}_1\psi_{1x}\psi_2\bar{\psi}_2+\bar{\psi}_2\psi_{2x}\psi_1\bar{\psi}_1-\bar{\psi}_2\bar{\psi}_{1x}\psi_2\psi_1-\bar{\psi}_1\bar{\psi}_{2x}\psi_1\notag\\
&&\psi_2)+(\bar{\psi}_{1xxx}\psi_1-\bar{\psi}_1\psi_{1xxx}-\bar{\psi}_{1xx}\psi_{1x}+\bar{\psi}_{1x}\psi_{1xx})+(\bar{\psi}_{2xxx}\psi_2-\bar{\psi}_2\notag\\
&&\psi_{2xxx}-\bar{\psi}_{2xx}\psi_{2x}+\bar{\psi}_{2x}\psi_{2xx}).
\end{eqnarray}
Based on the zero-curvature formulation of $\tilde{F}$ and $\tilde{G}$, we have the fermionic fifth-order NLSE,
\begin{eqnarray}
&&i\psi_{1t}+4i(\psi_{1x}\psi_2)_{xx}\bar{\psi}_2+4i(\psi_{1x}\psi_2)_x\bar{\psi}_{2x}+2i(\bar{\psi}_2(\psi_1\psi_{2x})_x+\psi_1(\psi_{2x}\bar{\psi}_2)_x-(\psi_1\bar{\psi}_{2x})_x\notag\\
&&\psi_2)_x-i\psi_{1xxxxx}+i\psi_1((\bar{\psi}_{2x}\psi_2-\bar{\psi}_2\psi_{2x})_{xx}-2(\bar{\psi}_{2xx}\psi_{2x}-\bar{\psi}_{2x}\psi_{2xx}))+i((\psi_{1x}\bar{\psi}_2-\notag\\
&&\psi_1\bar{\psi}_{2x})_{xx}-2(\psi_{1xx}\bar{\psi}_{2x}-\psi_{1x}\bar{\psi}_{2xx}))\psi_2=0,\notag
\end{eqnarray}
\begin{eqnarray}\label{NLSE2}
&&i\psi_{2t}+4i(\psi_{2x}\psi_1)_{xx}\bar{\psi}_1+4i(\psi_{2x}\psi_1)_x\bar{\psi}_{1x}+2i(\bar{\psi}_1(\psi_2\psi_{1x})_x-\psi_2(\bar{\psi}_1\psi_{1x})_x-(\psi_2\bar{\psi}_{1x})_x\notag\\
&&\psi_1)_x-i\psi_{2xxxxx}+i\psi_2((\bar{\psi}_{1x}\psi_1-\bar{\psi}_1\psi_{1x})_{xx}-2(\bar{\psi}_{1xx}\psi_{1x}-\bar{\psi}_{1x}\psi_{1xx}))-i((\psi_{2x}\bar{\psi}_1-\notag\\
&&\psi_2\bar{\psi}_{1x})_{xx}-2(\psi_{2xx}\bar{\psi}_{1x}-\psi_{2x}\bar{\psi}_{1xx}))\psi_1=0.
\end{eqnarray}

\section{Summary and  Discussion}

In this study, we have discussed the fifth-order HS model with two types of constraints which are $S^2=S$ and $S^2=3S-2I$, respectively.  We construct their  gauge equivalent equations, which are the super and fermionic fifth-order NLSEs. It is well known there is a close relationship between the HS model and the Hubbard model which attracts lots of interest in physics. This study further demonstrates that the method in this paper can be also applied to establish other higher-order HS models and higher-order super and fermionic NLSE. For the applications of the generalized HS models and higher-order related NLSEs constructed in this paper, we shall try to do it in a near future.

\section{Acknowledgements}
This work is partially supported by National Natural Science Foundation of China
(Grant Nos. 11965014, 11605096 and 11601247), We thank Prof. Ke Wu
and Prof. Weizhong Zhao (CNU, China) for long-term encouragement and support.

\end{document}